# Crowding-induced Elongated Conformation of Urea-unfolded Apoazurin: Investigating the Role of Crowder Shape In Silico

Fabio C. Zegarra[&,1], Dirar Homouz[&,1,2,3], Andrei G. Gasic[1,3], Lucas Babel[1], Michael Kovermann[4,5*], Pernilla Wittung-Stafshede[6,*], Margaret S. Cheung[1,3,*]

& Equal contribution

1. Department of Physics, University of Houston, USA
2. Department of Physics, Khalifa University of Science and Technology, Abu Dhabi, UAE
3. Center for Theoretical Biological Physics, Rice University, USA
4. Department of Chemistry, Universität Konstanz, Germany
5. Research School Chemical Biology (KoRS-CB), Universität Konstanz, Germany
6. Biology and Biological Engineering Department, Chalmers University, Sweden

**CORRESPONDENCE**

*Dr. Margaret Cheung  mscheung@uh.edu

*Dr. Pernilla Wittung-Stafshede  pernilla.wittung@chalmers.se

*Dr. Michael Kovermann  michael.kovermann@uni-konstanz.de

**Abstract**

Here, we show by solution nuclear magnetic resonance measurements that the urea-unfolded protein apoazurin becomes elongated when the synthetic crowding agent dextran 20 is present, in contrast to the prediction from the macromolecular crowding effect based on the argument of volume exclusion. To explore the complex interactions beyond volume exclusion, we employed coarse-grained molecular dynamics simulations to explore the conformational ensemble of apoazurin in a box of monodisperse crowders under strong chemically denaturing conditions. The elongated conformation of unfolded apoazurin appears to result from the interplay of the effective attraction between the protein and crowders and the shape of the crowders. With a volume-conserving crowder model, we show that the crowder shape provides an anisotropic direction of the depletion force, in which a bundle of surrounding rod-like crowders stabilize an elongated conformation of unfolded apoazurin in the presence of effective attraction between the protein and crowders.

# I. INTRODUCTION

The cellular environment is highly crowded with different types of macromolecules that can occupy volumes up to 40 % of the total volume inside cells [1-3]. To properly understand how proteins fold and function *in vivo*, it is necessary to consider the volume exclusion from surrounding macromolecules, known as the macromolecular crowding effect [4], which influences the structures and dynamics of a biopolymer in this restricted space. To test the crowding effect *in vitro*, sugar-based, branched polymers such as Ficoll and dextran [5] are typically utilized as synthetic crowder agents to investigate the macromolecular crowding effect because they are considered to be chemically inactive agents in solution. Under the assumption that the dominant interaction between a crowding agent and a biopolymer is steric repulsion, the macromolecular crowding is shown to stabilize the conformation of compact biopolymers as shown in proteins [6-8], nucleic acids [9], and chromatins [10].

However, here we made the puzzling discovery by solution nuclear magnetic resonance (NMR) spectroscopy that our model protein, apoazurin, unfolded by a high urea concentration adopts more *elongated* conformations in the presence of dextran 20 than without a crowding agent (bulk). This is a deviation from a previous study from the Wittung-Stafshede group where they have shown that apoazurin increases in stability with synthetic crowders [11]. The thermodynamic stability of the two-state folding protein apoazurin increases with the increased volume fraction of crowders, and this effect was independent of the crowder size when various dextran sizes were analyzed, suggesting that a cylindrical representation from Minton's model [12] is appropriate to model dextran that interacts with apoazurin through steric repulsions [13].

To find an explanation to this counterintuitive result, we considered *in silico* coarse-grained molecular dynamics simulations, allowing more complex modeling of crowders than a simple hard sphere model for volume exclusion. The question central to all computational modeling is how to choose an appropriate level of the coarse graining of crowders that captures essential features without a prohibitive computational cost. All-atom molecular dynamics simulations have shown an unprecedented resolution of protein and protein crowders that allows investigation of subtle chemical interactions affecting protein stability and kinetics [14-15]. These simulations demand massive computational resources and still require advanced sampling techniques to explore a wide range of phase space in a system for the computation of thermodynamic properties [16-17]. Coarse-grained models, although losing atomistic details, allow for greater sampling than in all-atom models and provide key polymeric insight into the distribution of protein conformations and the folding energy landscape under the macromolecular crowding effect [18-21].

To investigate the structural ensemble of apoazurin in 6 M urea in a periodic box of dextran crowders with coarse-grained molecular simulations, we must consider the appropriate dextran model. It is challenging to determine the geometry and property of dextran in a coarse-grained model because of its branched topology and polydisperse spatial distribution in solution [22]. Early experiments based on filtration have characterized that the shapes of dextran crowders are asymmetrical by showing that dextrans pass through a pore much smaller than their presumed diameters [5]. Sasahara *et al* have further modeled the shape of Dextrans as rods to fit the data from experimental measurement of the protein stability in dextrans with circular dichroism[12]. Our previous work has adopted a spherocylindral model for dextrans to explain the features of folding from various protein mutants[23]. Nevertheless, these studies have assumed dextrans to interact with a protein through hard-core interactions. Interestingly, studies by Jiao *et al.* [24] and, recently, by Ebbinghaus *et al.* [25] showed that dextran stabilizes a protein through enthalpic interactions similar to its building block glucose, instead of stabilizing through volume exclusion. A suitable choice of dextran models again has become elusive from the past literature.

A mixture of dextran with destabilizing urea complicates our analysis of apoazuring folding. The net opposing impact from dextran and urea on protein stability can depend on the composition and become nonadditive [26]. Given only the hydrodynamic radius of dextran in 6.8 M urea from the NMR experiments, our modeling approach was to vary the shape of a volume-conserving crowder model for dextran while changing the effective attraction between urea-unfolded apoazurin and its surrounding dextran crowders.

We first showed that a naïve crowder model of hard-core repulsion between dextrans and apoazurin fails to capture the experimental observation. The addition of an effective attraction between dextrans and apoazurin is necessary to shift the protein ensemble towards elongated conformations. These effective attractions are unlikely the ones such as electrostatics and hydrogen bonding interactions found in protein crowders [27-28]. We are attracted to another possible explanation that compressible (physically soft) crowders[29] or chemically soft repulsion[30] produce effective attractions between a protein and a crowder because of the increased depletion between soft repulsive depletants. It is noteworthy that the soft repulsion from Zaccarelli's work[29] is due to deformation of polymers, which is an entropic effect. In Harries's work[30-31], however, the soft repulsion from cosolute-solvent mixtures, resulting from changes in the hydrogen bonding in solvent near the first hydration shell, is represented by a step-like function in the interacting potential – an enthalpic effect. Our effective attractive potential can rise from either or both factors.

The geometry of an elongated rod model for dextrans further shifts the ensemble of apoazurin towards elongated conformations (favoring urea-unfolded conformations). The attraction between urea-unfolded apoazurin and dextrans facilitates dextrans to form aligned bundled conformations. The shape of the dextran breaks the symmetry of the depletion force, thus favoring an elongated geometry for the unfolded apoazurin. With coarse-grained modeling, we provide a plausible explanation to the puzzling discovery from the NMR experiments.

## II. MATERIALS AND METHODS

### II.1 Apoazurin

We here used the apo-form *Pseudomonas aeruginosa* azurin, termed apoazurin, to study the excluded-volume effects of dextran 20 on the urea-denatured protein *in vitro* and *in silico*. It is a bacterial protein that binds a copper ion that can undergo oxidation-reduction as part of electron transfer chains in bacteria. Apoazurin has 128 residues, and the folded state adopts a β-sheet Greek-key fold with a disulfide bond (Cys3-Cys26) [32] (see Figure 1). *In vitro* studies on apoazurin have advanced our knowledge on the folding mechanism of sandwich-like proteins [33]. Apoazurin has provided an excellent template to compare experimental measurement with the predictions from protein folding theories [34] and simulations [35].

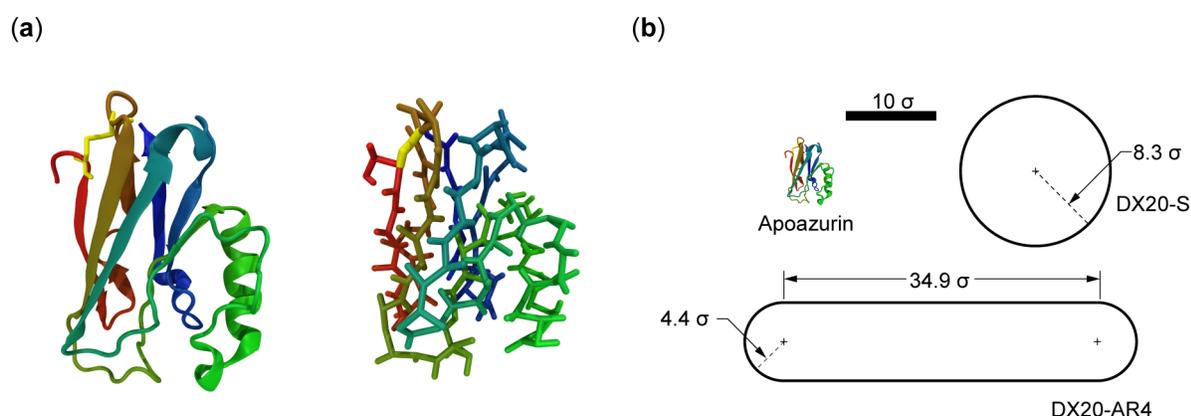

**Figure 1**. (a) Representations of the folded structure of apoazurin (PDB ID: 1E65). The folded structure is illustrated in a cartoon representation and a side-chain $C_\alpha$ representation. These protein illustrations were produced using VMD [36]. The algorithm DSSP [37] was used to assign the secondary structures. The residues that form the disulfide bond are shown in yellow. (b) Illustrations of two models for dextran 20 with the same volume: a spherical model (DX20-S), and a spherocylinder model with an aspect ratio of 4 (DX20-AR4). As reference, the folded structure of apoazurin is shown. The reduced unit of length is σ.

### II.2 NMR measurements

All NMR experiments were conducted at a Bruker Avance III HD 850 MHz spectrometer equipped with a z-gradient cryogenic probe using 20 mM NaP, pH 7.0 and 5 % $D_2O$ (v/v) as sample buffer. All one-dimensional data sets were processed and analysed by Topspin 3.2 whereas two-dimensional data sets were processed with NMRPipe [38] and analysed in NMRView [39]. Water suppression was achieved by presaturation and WATERGATE.

*II.2.1 Diffusion experiments.* NMR diffusion spectra were recorded using pulsed field bipolar gradient stimulated echo experiments [40] at $T = 298$ K. For each diffusion profile, 21 different gradient strengths $G$ were used for 6 ms along the z axis followed by a 100 ms recovery delay. The diffusion of apoazurin was allowed to proceed for 100 ms ($c^{d20} = 0$ g/l, $c^{d20} = 100$ g/l, $c^{d20} = 150$ g/l, $c^{d20} = 180$ g/l, $c^{d20} = 200$ g/l and $c^{urea} = 0$ M; $c^{d20} = 0$ g/l, $c^{d20} = 100$ g/l and $c^{urea} = 6.8$ M), 150 ms ($c^{d20}$ = 150 g/l and $c^{urea}$ = 6.8 M) or 200 ms ($c^{d20}$ = 180 g/l and $c^{urea}$ = 6.8 M). The calibration of $G$ was performed by a standard protocol [41]. For error estimation, four different gradient strengths were repeated (relative gradient strengths of 1, 10, 40 and 70 %). The measured $^1$H NMR spectra were integrated within the aliphatic signal region $I$ ( 0.5... 1.5 ppm) and fitted to Equation (1):

$$I(G)=I(0)\exp(-G^2\gamma^2\delta^2 D(\Delta-\delta/3)) \qquad (1)$$

where $\gamma$ is the gyromagnetic ratio, $\delta$ is the gradient length, $\Delta$ is the diffusion time and $D$ is the calculated diffusion coefficient [40].

*II.2.2 Determination of viscosity.* The viscosity η of the solvent has been determined by measurement of the diffusion coefficient $D$ of the internal reference molecules dioxane present at different conditions ($c^{d20}$ = 0 g/l and $c^{urea}$ = 0 M; $c^{d20}$ = 0 g/l and $c^{urea}$ = 6.8 M) or dextran 20 ($c^{d20}$ = 100 g/l, $c^{d20}$ = 150 g/l, $c^{d20}$ = 180 g/l, $c^{d20}$ = 200 g/l and $c^{urea}$ = 0 M; $c^{d20}$ = 100 g/l, $c^{d20}$ = 150 g/l, $c^{d20}$ = 180 g/l and $c^{urea}$ = 6.8 M), known hydrodynamic radii of dioxane, $r_H^{dioxane}$ = 1.87 Å [42] and $r_H^{d20}$ = 32.4 Å [43] and by applying Equation (2)

$$D = k_B T/(6\pi\eta r_H) \qquad (2)$$

with Boltzmann's constant $k_B$ and temperature $T$.

## II.3 Coarse-grained models

We have performed coarse-grained molecular simulations of apoazurin under several crowding conditions and under 0 M and 6 M of urea denaturant. 6 M urea is a concentration that guarantees the denaturation of the protein at room temperature.

*II.3.1 Coarse-grained apoazurin model.* Our model protein apoazurin (PDB ID: 1E65; Figure 1) is studied employing a coarse-grained model. We coarse-grained each residue into a side-chain and $C_\alpha$ bead with exception to glycine, which does not contain a side-chain [44]. The $C_\alpha$ bead is located at the position of the α-carbon, and the side-chain bead is located at the center of mass of the side-chain atoms. We have employed a structure-based Hamiltonian for our protein model, i.e., the energy landscape of the protein follows the principle of minimal frustration [45-47], by following previous work [44, 48]. The dimensions of the protein are expressed in units of σ (σ = 3.8 Å), which is the average distance between two consecutive $C_\alpha$ beads in a protein. This reduced unit of length is also used for the dimensions of the systems with crowders.

*II.3.2 Coarse-grained Dextran 20 model.* We employed twelve volume-conserving models on dextran 20, a glucose-based polydisperse polymer of 20 kDa molar mass, to investigate the effects of crowder shapes and effective attraction with a protein (Figure 2). The radius of $r_H^{d20}$ = 32.4 Å was taken from the current experimental measurement. The rod-like shapes used by the models are spherocylinders [21] with various aspect ratios (*AR*). A spherocylindral model is created by connecting several overlapping spheres where the center-to-center distance between adjacent spheres is the radius of a sphere. The *AR* of a spherocylinder is defined as the ratio *L/D*, where *L* is the distance between the centers of the two spheres located at the ends of the spherocylinder, and *D* is the diameter of a sphere. For hard-core crowder models, the *AR* varies from 0 (DX20-S), to 1 (DX20-AR1), 2 (DX20-AR2), 4 (DX20-AR4), 8 (DX20-AR8), and 16 (DX20-AR16), while keeping the volume the same as DX20-S with the radius of $r_H^{d20}$ = 32.4 Å which is equivalent to 8.3 σ. This value corresponds to the hydrodynamics radius of dextran 20 from our *in vitro* experiment, and it is approximately twice the radius of gyration $R_g$ of the folded state ($R_g^N \approx 4$ σ).

The crowders with DX20-series interact with a protein through repulsion, and the crowders with DX20A-series interact with the protein through attractive interactions. Table 1 shows the number of beads and radii of those beads needed to create the various *AR*s.

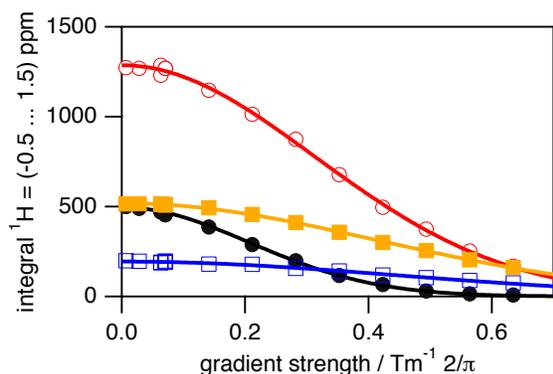

**Figure 2**. PFG-NMR based diffusion profiles obtained for native apoazurin in absence (colored in black, closed circles) and in presence of $c^{d20}$ = 100 g/l (colored in red, open circles), $c^{d20}$ = 150 g/l (colored in orange, closed rectangles) and $c^{d20}$ = 200 g/l (colored in blue, open rectangles). The straight lines are fits of Equation (1) to the experimental data. All data were recorded at $T$ = 298 K. Particular gradient strengths have been repeated (1, 10, 40, and 70 % related to maximum gradient strength possible).

| Name | Aspect ratio, *AR* | Number of beads | Radii of beads ($\sigma$) |
|---|---|---|---|
| DX20-S and DX20A-S | 0 | 1 | 8.3 |
| DX20-AR1 and DX20A-AR1 | 1 | 3 | 6.1 |
| DX20-AR2 and DX20A-AR2 | 2 | 5 | 5.3 |
| DX20-AR4 and DX20A-AR4 | 4 | 9 | 4.4 |
| DX20-AR8 and DX20A-AR8 | 8 | 17 | 3.6 |
| DX20-AR16 and DX20A-AR16 | 16 | 33 | 2.9 |

**Table 1** Physical characteristics of the coarse-grained model of dextran 20. The spherocylinders are characterized by aspect ratio *AR*, number of beads needed to compose the spherocylinder, and the radii of those beads. DX20-series are crowders with repulsive interactions with a protein. DX20A-series are crowders with attractive interactions with a protein.

An apoazurin protein is placed in the center of the box of crowders of the same type. The number of the crowders in a box depends on the volume fraction $\phi_c$ under study. The volume fractions studied are $\phi_c$ = 0 % (or bulk), 5 %, 10 %, and 20 %.

*II.3.3. The energy function of apoazurin under urea interference in a periodic cubic box of crowders.* The Hamiltonian of the system ($E_s$) is the sum of the potential energy of the protein ($E_p$), the interaction between protein and crowders ($E_{pc}$), and the interaction between crowders ($E_{cc}$), i.e., $E_s = E_p+E_{cc}+E_{pc}$. $E_p$ adopts a variant of structure-based potentials where only the contacts found in the native states are attractive according to the amino acid types, while the non-native contacts are repulsive. We have incorporated the chemical interference by adopting the "Urea-SC-DIH" model from a previous study [48] where the urea solvent-mediated interactions were included in the interactions between side-chain beads and the dihedral-potential that corresponds to four successive $C_\alpha$ beads. The expression of $E_p$ can be found in the supplementary information.

For $E_{cc}$ the interactions between crowders are repulsive and inversely related to the distance between the two crowders to the 12$^{th}$ power [Equation (S3)]. We set the interactions between protein and crowders, $E_{pc,}$ to be repulsive or attractive. The expression of repulsion between hard spheres of crowders labeled as DX20-series in Table 1 follows the same expression as $E_{cc}$. The expression of non-specific attraction between a crowder labeled as DX20A-series in Table 1 and a bead from the apoazurin model follows a Lennard-Jones potential with the depth well of $\varepsilon_s$. The relative strength of the attractive interaction is denoted by $\lambda = \varepsilon_s/\varepsilon$, where $\varepsilon = 0.6$ kcal/mol is the solvent-mediated interaction between Thr and Thr residues at 0 M. We used $\lambda = 0.83$ and $2.50$ for this study.

## II.4. Molecular simulations

We used an in-house version of the Amber 10 molecular dynamics package [49] to integrate the Langevin equation of motion in the low friction limit [50]. The integration time step was set to $10^{-4}\tau_\varepsilon$, where $\tau_\varepsilon = \sqrt{(m\sigma^2/\varepsilon)}$, $m$ is the mass of a $C_\alpha$ bead, $\varepsilon$ is the solvent-mediated interaction, and $\sigma$ is the van der Waals radius of a $C_\alpha$ bead. The replica exchange method (REM) [51-52] was used in order to efficiently sample the configurational space of samples when required. The exchanges between the neighboring replicas $i$ and $j$ are attempted every 40 $\tau_\varepsilon$. The acceptance or rejection of each exchange follows the Metropolis criterion, $min\{1, exp[(\beta_i - \beta_j) \cdot (U(r_i) - U(r_j))]\}$, where $\beta = 1/k_B T$, $k_B$ is the Boltzmann constant, $T$ is the temperature, and $U(r)$ is the potential energy of the system. Twenty replicas were employed where the temperature range used in a REM simulation varies for each condition. For each replica, more than 100 000 statistically significant conformations were collected.

For each condition of the simulations with crowders, there are ten sets of randomly assigned crowder configurations in a periodic boundary cubic box (PBC) as initial conditions. The size of PBC is $120\times120\times120$ $\sigma^3$ for the crowders with $AR$ from 0 to 8, and it is $185\times185\times185$ $\sigma^3$ for the crowders with $AR = 16$. The canonical simulations were performed at a temperature that corresponds to the *in silico* room temperature $T_u$. $T_u$ is obtained by transferring the difference between the melting temperature ($T_m$) and the room temperature (298 K) for an *in vitro* experiment to the *in silico* computational study as suggested in a prior study [35]. The *in silico* room temperature is given by

$$T_u(in\ silico) = T_f(in\ silico) - [T_m(in\ vitro) - 298\ K]. \qquad (3)$$

Where $T_f$ is the *in silico* folding temperature. In our case the *in silico* room temperature at 0 M urea is $k_B T_u/\varepsilon = 1.14$.

## II.5 Data Analysis

*II.4.1 Geometrical properties of a protein.* We measured the geometry of apoazurin by using the radius of gyration ($R_g$) and asphericity parameter ($\Delta$) [53] [see Equation (S7)]. $R_g$ measures the size of the protein, and $\Delta$ measures the degree of anisotropy of the structure of the protein. $\Delta$ has the following range: $0 \leq \Delta \leq 1$, where a perfect sphere has $\Delta = 0$. When $\Delta$ approaches 1, it is an indication that the protein is fully extended.

*II.4.2 The geometrical properties of a crowder void.* The crowder void is defined as the space between the crowders in which the protein lies. To find this void, we partition the entire simulation box through Voronoi tessellation [54]. From the tessellation, a 3D polygon is created around each bead (both crowders and protein residues). We select the crowder beads whose polygon is in contact with the polygon of the protein beads, and the positions of these selected crowder beads are used to define the vertices of the crowder void. We characterize the crowder void by computing two geometrical descriptors: the radius of gyration $R_g$ and the asphericity parameter $\Delta$.

## III. RESULTS

### III.1. Probing apoazurin *in vitro*

*III.1.1. Structural changes*

It has been established that high-resolution NMR spectroscopy enables the profound molecular characterization of the structural [55] and dynamic properties [56] of a protein ensemble as it is present in solution. We have made use of that by applying two-dimensional heteronuclear $^1$H-$^{15}$N NMR experiments on apoazurin present under native and denatured conditions in a macromolecular crowding environment. Moreover, we have used Pulse-field gradient (PFG)-NMR techniques to decode the potential impact of macromolecular crowding on the hydrodynamic dimension of apoazurin present under both folded and denatured conditions.

    The addition of dextran 20 to apoazurin up to a concentration of $c^{d20} = 200$ g/l neither changes the chemical shifts nor the signal intensities of apoazurin as probed by one-dimensional proton (Figure S1) as well as by two-dimensional $^1$H-$^{15}$N NMR HSQC spectroscopy (Figure S2). This indicates that the addition of dextran 20 does not modify the structural properties of folded apoazurin at secondary and tertiary levels.

    The maximum change in $^1$H, $^{15}$N weighted chemical shifts, $\Delta\omega_{max}$, is 0.025 ppm with an average $\Delta\omega^{mean}$ of (0.012±0.004) ppm. Note that chemical shifts for aliphatic protons of apoazurin are highly conserved as well by comparing diluted and $c^{d20} = 200$ g/l conditions for the folded state. Next, dextran 20 of up to $c^{d20} = 180$ g/l was additionally supplemented to apoazurin that has been chemically unfolded using $c = 6.8$ M of urea (Figure S3 and S4).

Again, changes in $^1$H, $^{15}$N weighted chemical shifts are small, with $\Delta\omega_{max} = 0.025$ ppm and $\Delta\omega^{mean} = (0.009\pm0.006)$ ppm. Chemical shifts for aliphatic protons are conserved in the presence of $c^{urea} = 6.8$ M by comparing diluted and $c^{d20} = 180$ g/l conditions as observed for folded apoazurin.

Taken together, 200 g/l of dextran 20 does not change the structural integrity of the folded state and 180 g/l of dextran 20 does not perturb the chemical shift pattern seen for chemically unfolded apoazurin. Consequently, we conclude that there is no significant solute-protein interaction between dextran 20 and apoazurin.

### *III.1.2. Molecular dimension*

NMR diffusion methodology has been successfully employed for the profound characterization of the molecular dimensions of both folded and highly denatured proteins in solution [42]. The initial idea with these experiments was to probe the expected (based on excluded volume theory) compaction of unfolded apoazurin when macromolecular crowding agents have been added.

First, we probed the diffusion properties of apoazurin in the absence of urea. The diffusion coefficient, $D$, obtained for folded apoazurin ranges from $D = (11.4\pm0.3)\times10^{-11}$ m$^2$s$^{-1}$ for diluted conditions to $D = (2.44\pm0.07)\times10^{-11}$ m$^2$s$^{-1}$ present at $c^{d20} = 200$ g/l (Figure 2).

As the solvent viscosity $\eta$ increases from $\eta = 0.89\pm0.01$ mPa·s to $\eta = 3.99\pm0.06$ mPa·s (Figure 3), the hydrodynamic radius $r_H$ for folded apoazurin remains constant at $r_H = 21.7\pm0.8$ Å. Notably, $r_H$ determined for folded apoazurin is in excellent agreement at all concentrations of dextran 20 with the molecular dimension expected for a globular protein of azurin's size [42]—i.e., $r_H(N^{aa}=128) = (20\pm5$ Å) (Figure 4).

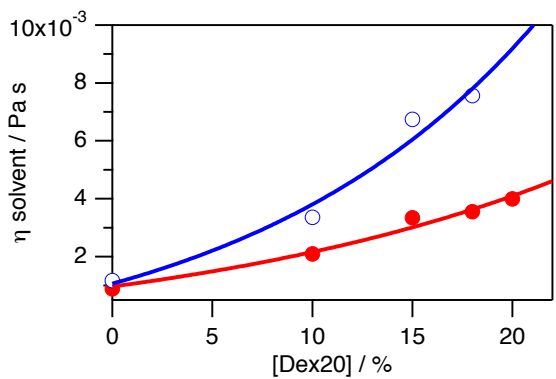

**Figure 3**. The viscosity η of the solvent depends on the concentrations of dextran 20 and urea (absence of urea: closed circles, colored in red; presence of $c$ = 6.8 M urea: open circles, colored in blue) and has been calculated by determining the diffusion of dioxane and dextran 20 and Equation (2) at $T$ = 298 K. The straight lines have been computed by assuming a polynomial function to guide the eye. On page 11 in the caption of Figure 3:… Error bars have been included. They are too small to be seen. They have been calculated based on the standard deviation of the diffusion coefficient obtained by fitting Eq. 1 to the diffusion data. This error of the diffusion coefficient has been used to determine the error of rH by using Eq. 2.

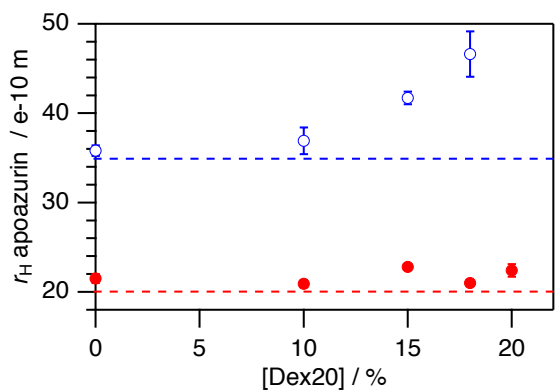

**Figure 4**. The hydrodynamic radius $r_H$ of apoazurin depends on the macromolecular crowding conditions used. Whereas native apoazurin does not show a dependence on the concentration of dextran 20 (closed circles, colored in red), denatured apoazurin increases in the hydrodynamic dimension by increasing the concentration of dextran 20 (open circles, colored in blue). The horizontal lines (dotted mode) show the hydrodynamic radii that are theoretically expected for a globular folded protein (colored in red) and a denatured polypeptide chain of that size (colored in blue) according to [42]. All the data were recorded at $T$ = 298 K. Error bars have been included (and seen). The calculation has been done in the same way as for we did Figure 3.

Next, we probed the diffusion coefficient for apoazurin under denaturing conditions of $c^{urea}$

= 6.8 M. Again, $D$ decreases with an increasing concentration of dextran 20 from $D$ = $(5.22\pm0.08)$e-11 m$^2$s$^{-1}$ ($c^{d20}$ = 0 g/l) to $D$ = $(0.62\pm0.06)$e-11 m$^2$s$^{-1}$ ($c^{d20}$ = 180 g/l). However, the decrease in $D$ was not compensated by the increase in $\eta$ as found under native conditions. Instead, surprisingly, the hydrodynamic radius for denatured apoazurin *increases* from $r_H$ = 35.8$\pm$0.6 Å ($c^{d20}$ = 0 g/l) to $r_H$ = 46.6$\pm$2.5 Å ($c^{d20}$ = 180 g/l) (Figure 4). Notably, $r_H$, as determined here for denatured apoazurin under dilute conditions, agrees with the molecular dimensions expected for a denatured polypeptide chain of apoazurin's size [42], $r_H(N^{aa}=128)=(35\pm18$ Å). Because excluded volume theory predicts that unfolded proteins become compacted in the presence of macromolecular crowding, more factors must be at play. To find a molecular-mechanistic explanation to this *in vitro* observation, we turned to *in silico* computations.

### III.2. *In silico* analysis of apoazurin unfolding

To compare our simulation results to the experimental measurements of apoazurin's hydrodynamic radii $r_H$ under different conditions, we used two order parameters: radius of gyration $R_g$ and asphericity $\Delta$. $R_g$ is not an exact comparison to $r_H$ because the shape of the object also matters in determining $r_H$. For a perfect sphere, $r_H$ can be approximated by $R_g$; however, as the protein becomes less spherical (or an increase in $\Delta$), $r_H$ and $R_g$ diverge (see Figure S5). $r_H/R_g$ can exceed 3 when $\Delta$ approaches 1 (a rod-like shape). This is due to the diffusion of spherical molecules being different than that of other elongated molecules. Therefore, $r_H$ is in fact measuring a combination of the *size and shape* of a molecule.

#### III.2.1. Structural changes for the repulsive crowder model and chemical denaturation

We have investigated the ensemble conformations of apoazurin in various crowding contents at a chemically denatured state at 6 M urea and at an *in silico* room temperature of $k_B T_u/\varepsilon$ = 1.14 (see Methods) with coarse-grained molecular simulations.

First, we have assumed that crowders and apoazurin interact through a repulsive potential that accounts for the excluded volume effects at $\phi_c$ = 5 %, 10 %, and 20 %. While conserving the volume of the crowders, we tested the shape effect by varying the aspect ratio of the crowders on apoazurin unfolding. Figure 5(a) shows the average asphericity ($\Delta$) normalized by the asphericity ($\Delta_B$) under the bulk condition for DX20-S, DX20-AR4 and DX20-AR8 against $\phi_c$. $\Delta/\Delta_B$ are all less than 1 over a wide range of $\phi_c$, implying that all dextran 20 models with repulsive interactions with apoazurin do not contribute to the elongation of apoazurin. For most of the conditions, the macromolecular crowding effect leads to compaction of apoazurin with $R_g/R_{g,B}$ less than 1 [Figure 5(b)]. The crowders with a rod-like geometry show greater compaction than the spherical crowder over all $\phi_c$ because rod-like crowders render a larger covolume than spherical crowders. Because the effects found upon including only volume exclusion cannot explain the experimental results of dextran 20-induced elongation of unfolded apoazurin, we next investigated a dextran 20 model that also provides effective attractive interactions.

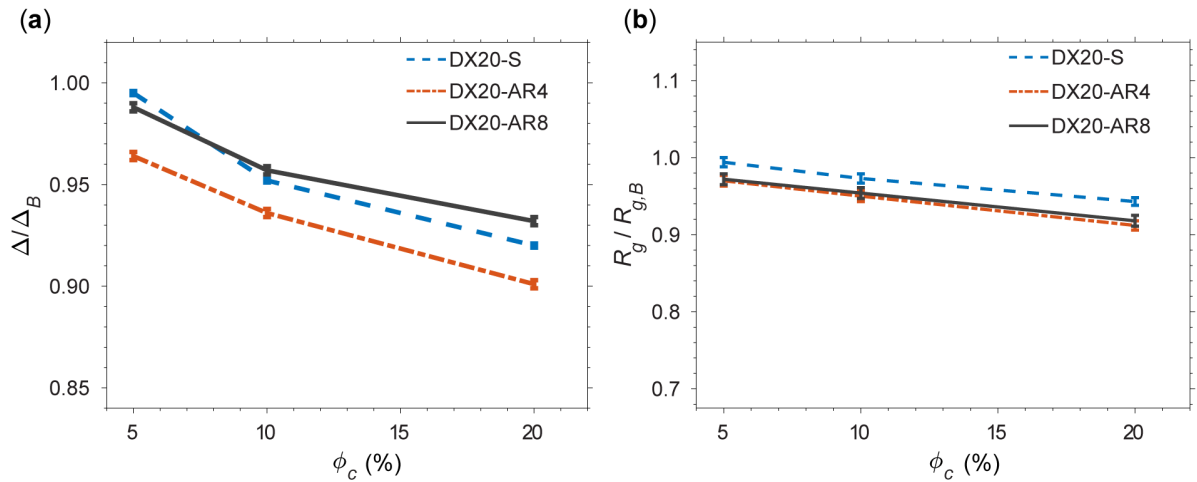

**Figure 5.** (a) Average asphericity ($\Delta$) of the protein apoazurin in varying shapes of crowders normalized by the average asphericity ($\Delta_B$) of apoazurin under the bulk condition against the volume fractions of crowders ($\phi_c$). (b) Average radius of gyration ($R_g$) of apoazurin normalized by the average radius of gyration under the bulk condition against $\phi_c$. The interactions between the protein apoazurin and crowders are repulsive. The concentration of urea is $c^{\text{urea}} = 6$ M. The temperature is $k_B T_u / \varepsilon = 1.14$, corresponding to room temperature (see Method), for all conditions. Error bars are included.

### *III.2.2. Structural changes of apoazurin with which crowders interact attractively and under chemical denaturation*

We hypothesize the existence of non-specific attractive interactions between the unfolded protein and crowders. For this, we tested two types of non-specific attractive interactions with $\lambda = 0.83$ and $\lambda = 2.50$ (see methods). $\lambda$ measures the attraction between a protein and crowder relative to the attraction between two Thr residues at 0 M urea. $\lambda = 0.83$ represents a weak attraction, and $\lambda = 2.50$ represents a strong attraction.

We noticed that a box of spherical crowders with a weak attraction (DX20A-S at $\lambda = 0.83$) cannot produce elongated ensembles of apoazurin *in silico* at room temperature with $k_B T_u / \varepsilon = 1.14$ in Figure 6. The distribution of $\Delta$ for apoazurin, centered at $\Delta \approx 0.16$, is significantly narrower than the bulk condition over a wide range of $\phi_c$. We note that, with a strong attraction at $\lambda = 2.5$, there was a slight increase in $\Delta/\Delta_B$ or $R_g/R_{g,B}$ along $\phi_c$ (Figure S6). However, $R_g/R_{g,B}$ is still less than 1 when the aspect ratio of the crowder is less than 4.

This result leads us to a working hypothesis that, in addition to weak attractions, the crowders must be rod-like to stabilize an extended unfolded conformation of apoazurin. We showed that, in the system of weakly attractive crowders with a rod-like shape (DX20A-AR4), the distribution of $\Delta$ for the protein apoazurin under the chemically denatured condition becomes broad and comparable to that under the bulk condition in Figure 6. Interestingly, the distribution of $\Delta$ for apoazurin shifts toward $\Delta = 1$ and forms a large peak at $\Delta \approx 0.84$ at $\phi_c = 5$ % for DX20A-AR8 (Figure 7). This finding implies that more elongated conformations of apoazurin in chemically unfolded states are sampled in a system with the crowder model DX20A-AR8 than with the crowder model DX20A-AR4. However, the height and position of the peak shifts toward $\Delta = 0$ with increasing $\phi_c$. At a high $\phi_c$ of 20 %, apoazurin adopts a compact configuration again (Figure 8). This feature was not observed for spherical crowders (DX20A-S) with a weakly attractive interaction with a protein at $\lambda = 0.83$ (Figure S6).

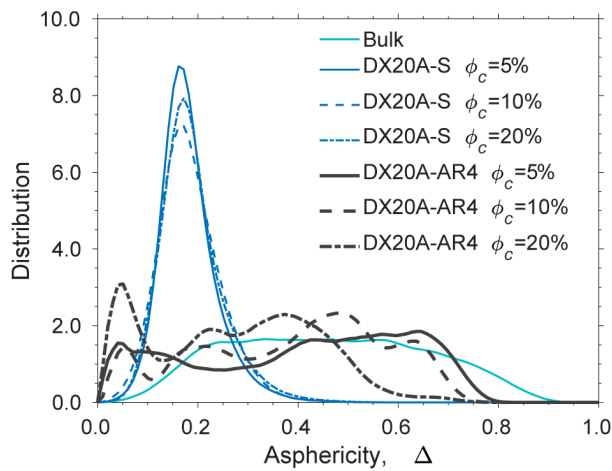

**Figure 6**. Distribution of the asphericity ($\Delta$) at $c^{\mathrm{urea}} = 6$ M in bulk, with DX20A-AR0, and with DX20A-AR4 with a strength of attractive interactions $\lambda = 0.83$ at a wide range of volume fraction of crowders ($\phi_c$). The temperature is $k_\mathrm{B} T_u / \varepsilon = 1.14$.

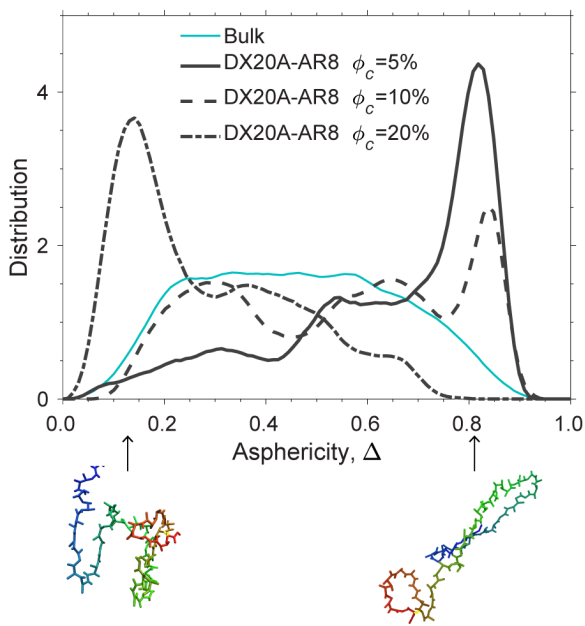

**Figure 7**. Distribution of the asphericity (Δ) for chemically unfolded apoazurin at 6 M urea in bulk and in a system of weakly attractive crowders (DX20A-AR8 at λ = 0.83) over several volume fractions ($\phi_c$). The temperature for all conditions is $k_B T_u/\varepsilon = 1.14$. Two representative structures are shown: at the left bottom with Δ = 0.14 ($\phi_c$ = 20 %) and the right bottom with Δ = 0.83 ($\phi_c$ = 5 %).

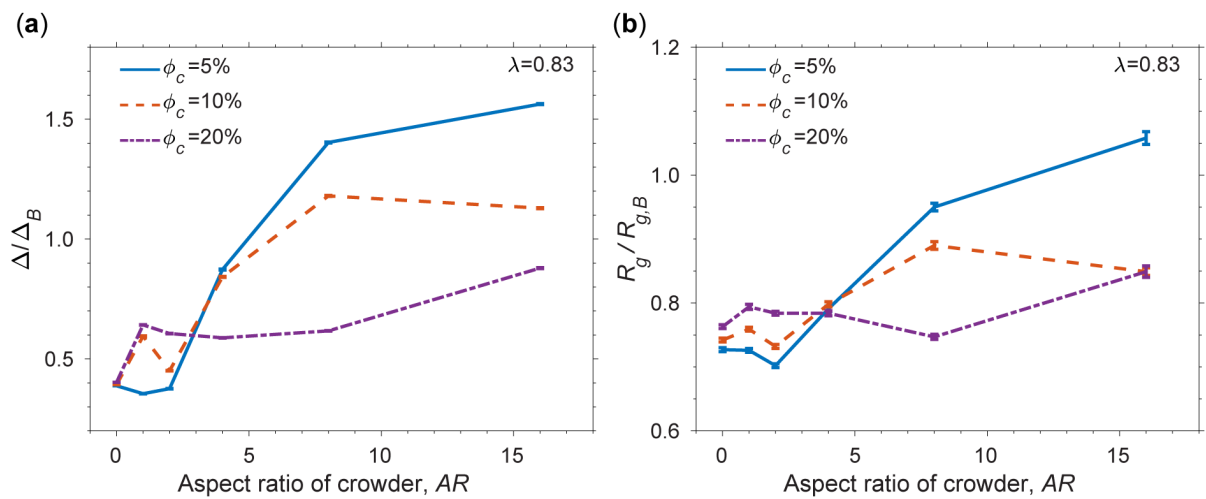

**Figure 8**. (a) Ratio of the average of the asphericity (Δ) under a specific condition and average of the asphericity ($\Delta_B$) under the bulk condition. (b) Ratio of the average of the radius of gyration ($R_g$) under a specific condition and average of the radius of gyration ($R_{g,B}$) under the bulk condition. These ratios are shown as a function of the aspect ratio of crowders, *AR*. The concentration of urea is 6 M, and the temperature is $k_B T_u/\varepsilon = 1.14$ for all conditions. Error bars are included.

*III.2.3. Both the rod-like shape of a crowder and attractive interaction with a protein contribute to the structural changes of apoazurin under chemically denatured conditions.*

We evaluated $\Delta/\Delta_B$ as well as $R_g/R_{g,B}$ of chemically denatured apoazurin at varying aspect ratios ($AR$) of crowders for a wide range of $\phi_c$ in Figure 8. Overall, there is an increase in $\Delta/\Delta_B$ when $AR$ increases [Figure 8(a)]. $\Delta/\Delta_B$ exceeds 1 at $AR$ greater than 8. Interestingly, $R_g/R_{g,B}$ also exceeds 1 at the most extreme case of $AR = 16$ [Figure 8(b)] at $\phi_c = 5$ %. However, we noted that the increasing trends in $\Delta/\Delta_B$ or $R_g/R_{g,B}$ are not uniform with $\phi_c$s. When $AR$ is less than 4, the protein has the largest $\Delta$ and $R_g$ values at $\phi_c = 20$ % compared with the other $\phi_c$ conditions. When $AR$ is greater than 4, the largest $\Delta$ and $R_g$ values occur at $\phi_c = 5$ %.

    Such a crossover behavior at $AR = 4$ is contributed by the shape of the protein-crowder assembly (Figure 9). The unfolded ensemble of apoazurin conformations adopts a shape according to the available void formed by surrounding crowders [Figure 9(a)]. A rod-like crowder with non-specific attractive interactions provides an elongated surface where several regions of the protein can stick [Figure 9(b)]. This particular geometry of rod-like crowder favors the formation of extended unfolded conformations of apoazurin. As several crowders stick to a protein, they tend to align themselves and restrict the available space for the protein. It appears that, at high $\phi_c$, there is insufficient space to accommodate the volume of a large assembly of protein crowders. Therefore, the uptick of $\Delta/\Delta_B$ or $R_g/R_{g,B}$ at high $\phi_c$ is weaker than that at low $\phi_c$.

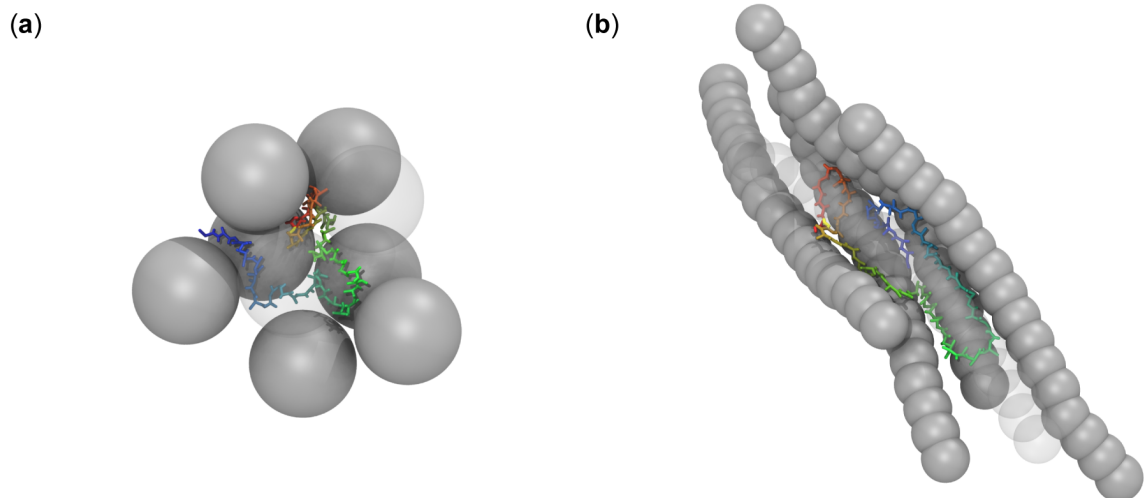

**Figure 9**. Snapshots of an exemplarily unfolded structure of apoazurin and the void between the protein and closest crowders for (a) spherical crowders (DX20A-S) (b) and rod-like crowders (DX20A-AR8) at $\phi_c = 5$ % and at $c^{urea} = 6$ M with weakly attractive interactions ($\lambda = 0.83$).

*III.2.3. The geometry of crowder-induced voids is modulated by the protein-crowder interactions*

Next, we investigated the geometry of voids, in terms of shape and size, formed by crowders of the various models modulated by the protein-crowder interactions. The voids are the depletion zones due to non-overlapping volume exclusion between a protein and crowders. The distribution of the asphericity of the void $\Delta^{void}$ is virtually unchanged between spherical and rod-like crowders without attractive interactions [Figure 10(a)]. Additionally, as $\phi_c$ increases, the peaks of the distributions are unperturbed. These results indicate that, without attractive interactions, the variation in the shape effect from rod-like crowders has been averaged in the ensemble. However, when attractive interactions are added to the crowders, there is a clear shift rightward (towards higher asphericity) of the peak. The spread of the distributions increases with $AR$ when the shape of a crowder becomes elongated [Figure 10(b)]. These effects are greatest for $\phi_c = 5\ \%$.

We show that the distribution of the size of a void ($R_g^{void}$) depends on the volume fraction of crowders, the shape of a crowder, and interaction between the protein and crowder. To make a fair comparison between spherical and rod-like crowders, we plotted the distribution of ($R_g^{void} - R$)/$\sigma$ in Figure 10(c) and Figure 10(d), where $R$ is the radius of a sphere in a crowder model. The overall size of ($R_g^{void} - R$)/$\sigma$ is smaller for rod-like crowders (DX20-AR8) than spherical crowders (DX20-S) by roughly a third at each $\phi_c$ when a protein and crowders interact through hard-core interaction [Figure 10(c)]. However, with a weak attraction between a protein and crowders ($\lambda = 0.83$), ($R_g^{void} - R$)/$\sigma$ is reduced by half across all crowder types [Figure 10(d)] compared to those in Figure 10(c). The peaks in the distribution of ($R_g^{void} - R$)/$\sigma$ for rod-like crowders (DX20A-AR8) at all $\phi_c$ nearly overlap at the same position around 10. This feature is consistent with the finding of a protein-crowder assembly in Figure 9.

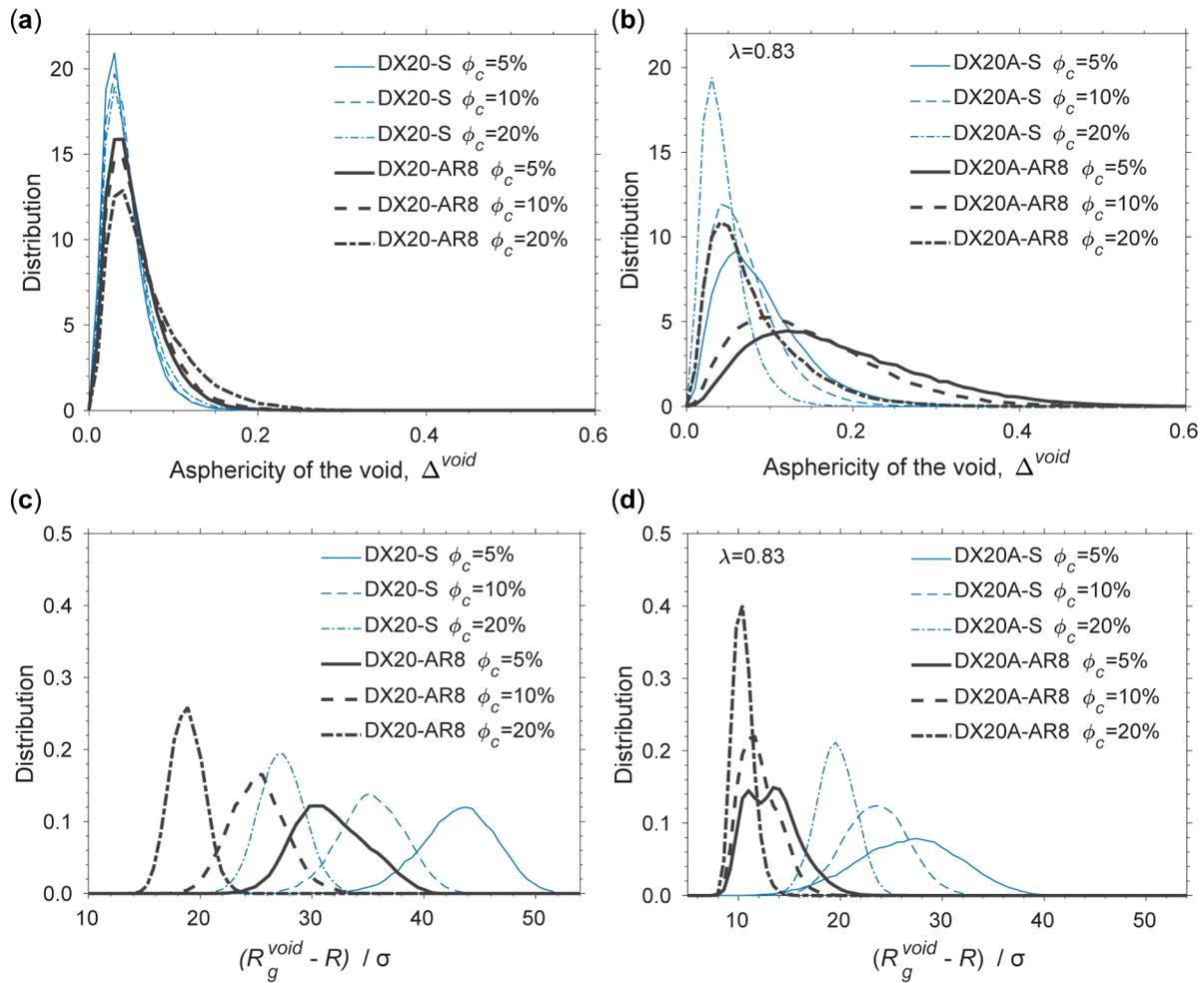

**Figure 10**. Distribution of the asphericity of the void ($\Delta^{void}$) between the protein and crowders with (a) steric repulsive interactions and (b) non-specific attractive interactions between the protein and crowders. The distribution of the radius of gyration of the void ($R_g^{void}$-R)/σ between the protein and crowders with (c) steric repulsive interactions and (d) weakly attractive interactions between the protein and crowders.

## IV. DISCUSSION

This *in silico* investigation was initiated by the anomalous observation from NMR experiments that the addition of 180 g/l of dextran 20 alters the ensemble of urea-unfolded apoazurin in the direction of an extended conformation. Initially, this observation contrasts the physical intuition of the excluded volume theory that proteins generally favor a collapsed, compact conformation in the presence of macromolecular crowders [57]. This intuition is built on a hard-sphere crowder (or depletant) that exerts entropic depletion interactions, or the macromolecular crowding effect [4], on the protein of interest, that effectively makes the unfolded expanded states of a protein entropically unfavorable, thus favoring a collapsed, compact conformation over an unfolded, extended one [18]. The strength of the macromolecular crowding effect grows with the volume fraction of crowders and is often strongest when the radius of the crowders is smaller than the $R_g$ of the test protein [58]. However, recent studies have shown that the sequence-dependent properties of a protein from protein crowders can offset the volume exclusion of protein crowders [24, 59]. The sequence dependent interactions between a protein and protein crowder include chemically soft interactions such as hydrogen bonding, electrostatic interactions, or dispersion. In some cases, these chemical attractions that possibly stick a protein to the surface of crowders further form quinary structures and destabilize the native folded state [28][60-61].

Chemically inert crowders such as Ficoll 70, PEG, dextrans do not share the same chemical surfaces as protein crowders. However, a spherocylinder model for dextrans that lacks atomistic details and interacts with a protein through hard-core potentials fails to explain the anomaly from experiments, unless we consider the effective attractions between a protein and dextrans. A recent study by Zhou *et al.* also suggests that synthetic crowders may attain strong attractions with intrinsically disordered proteins (IDPs) such that IDP polypeptides populate expanded configurations under crowded conditions [62]. The underlying reasons may include soft repulsion between the protein and synthetic crowders, deepening the depletion in the potential of mean force [29-30]. This physically soft aspect of an interaction (in contrast to chemically soft as mentioned above) is due to the compressibility of soft-matter materials that does not exist in non-overlapping hard spheres. It is possible that, in the mixture of urea and dextran, dextran becomes physically soft. This speculation is supported by a recent essay from Ferreira *et al.* [26] that, in the mixture of osmolyte and a macromolecular crowder, the solvent properties may not always be additive, depending on the composition of the mixture. We speculate that, at 6 M urea, not only the solvent-mediated interactions between amino acids in the protein are weakened, leading to unfolding, but dextrans also become softened to trap unfolded apoazurin conformations. This scenario can be imagined as spaghetti trapped by a fork. This "fork-spaghetti" mechanism may be facilitated by the non-uniform length of the branches of dextran [63]. The geometric feature of branched dextrans is at odds with the dimensions of a linear dextran model [12], especially when the molecular weight increases [22]. Noteworthily, the behavior of branched polymers (dextran) is distinct from those of linear polymers (PEG) with a similar molecular weight shown by Soranno *et al.* [64] who investigates the macromolecular crowding effect on the compaction of intrinsically disorder peptides (IDPs). When large PEG polymers reach a semi dilute region, they unwind and overlap one another at a moderate volume fraction of 10%. The monomers of overlapping PEG polymers still exert volume exclusion effect on the compaction of intrinsically disorder peptides (IDPs). However, the strength of volume exclusion from PEG monomers are not as strong as non-overlapping PEG polymers as predicted from the scaled particle theory [57].

Notably, we showed that the effective attraction alone is not sufficient to bias the ensemble of unfolded apoazurin towards an elongated one as the shape of the crowder must be cylindrical with a sufficiently large aspect ratio. With spherical crowders containing either weak or strong effective attraction, the enhanced depletion interactions still favor a compact unfolded apoazurin configuration [Figure 8(b)]. This contrasts the findings from Zhou's work in which peptide model samples with both extended and compact conformations interacted strongly with spherical crowders [62]. We speculate that the biphasic distribution might change if the length of the simulation is increased or if a more efficient sampling scheme is introduced. Because both the aspherical shape of a crowder and its effective attraction with a protein increase the correlation time in the simulations, we ensured that our conclusions are independent of the sampling scheme in our simulations. Our results at a volume fraction of 5 % for the DX20A-AR8 model at $\lambda = 0.83$ do not depend on the initial unfolded conformation of apoazurin. The temporal evolution of $\Delta$ is shown in Figure 11 at the early stage of two canonical simulations: one starting from a compact unfolded conformation and the other from an extended unfolded conformation. In the first case, $\Delta$ fluctuates and rearranges initially, and then it increases as time increases. In the second case, the conformation becomes compact from a stretched one, and then it rearranges to a have a less extended shape. Thus, the initial conformation is not the reason why the simulations favor the formation of extended conformations with crowders and non-specific attractive interactions.

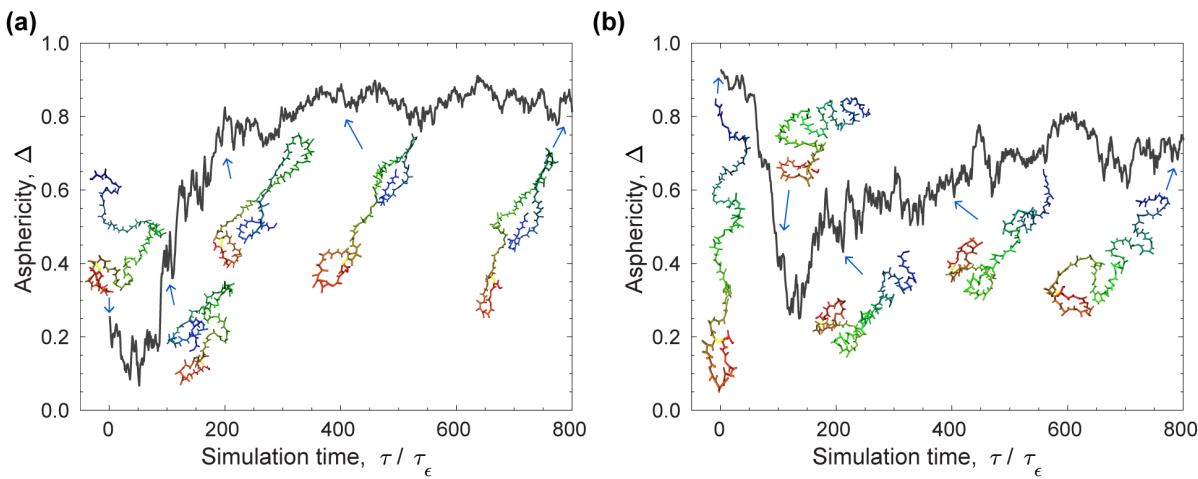

**Figure 11**. Initial stage of two canonical simulations for the model DX20A-AR8 at $k_BT_u/\varepsilon = 1.14$: one starting from (a) a compact configuration and the other from (b) an extended configuration. The snapshots of the protein depict the temporal structural changes.

To explain our *in silico* results, the elongated shape of a crowder must break the symmetry of the isotropic direction of the depletion force, creating an anisotropic force that sandwiches the backbone of the protein. We found that the elongated conformation of apoazurin is stabilized by a bundle of rod-like crowders in the presence of effective attraction. The rod-like crowders are still free to rotate, and there is no evidence of an isotropic-nematic transition in our system. In the absence of effective attraction, the repulsive cylinders do not self-assemble into nematic states. To reach a nematic state for rod-like crowders with an aspect ratio of 5, $\phi_c$ typically must exceed 40 % [65]. In fact, hard spherocylinders contribute to greater compactness of apoazurin because of an increased covolume compared with a volume-conserving hard sphere [4]. The difference here is the effective attraction of apoazurin and several rod-like crowders that collectively form a bundle. The bundles formed around elongated apoazurin maximize the volume accessible to the other crowders (not in a bundle), thus maximizing the total entropy of the system. We showed that elongated apoazurin appears at low $\phi_c$ = 5 %, but not at high $\phi_c$ = 20 %; the latter was observed in *in vitro* experiments. This difference may be due to the increase in excluded volume interactions between hard spherocylinders instead of the spherocylinder-apoazurin interactions, as the volume fraction of spherocylinders increases. To induce the formation of significant bundles at high $\phi_c$, the concentration of apoazurin in a binary mixture of spherocylinder and apoazurin should be increased.

## V. CONCLUSION

New understanding in science generally begins when theoretical expectations and new experimental findings do not agree. Here, the intriguing NMR discovery showing elongated urea-unfolded states in the presence of dextran goes against intuition from conventional macromolecular crowding theory and previous experimental findings. With very little clues from experimental data, we have employed coarse-grained modeling to investigate the underlying mechanism. Even though we cannot definitively exclude other possible explanations for the trends found in the experiments, our work denotes the importance of both crowder shapes and solvent-mediated interactions in manipulating the geometry of a protein. These two properties are pivotal to understand the full complexity of proteins in the cytoplasm. Furthermore, this newfound understanding can potentially be developed into a design principle for probes to accurately measure crowding volume fraction, shapes, and interactions locally in a subcellular environment.

**Supporting Information**
Hamiltonian of the coarse-grained protein model in the presence of urea, NMR measurements (Figure S1-S4), definition of the asphericity Δ, ratio of $r_H$ and $R_g$ as a function of Δ (Figure S5), changes in the geometry of the protein as a function of volume fraction of crowders (Figure S6), and protein stability at different concentrations of urea (Figure S7) are presented in the supporting information (SI).

**Acknowledgements**
We thank Alexander Christiansen for providing apoazurin and Birgit Köhn for technical assistance. We acknowledge the funding support from the National Science Foundation (MCB-1412532 and ACI-1531814) and the computational resources from the Center for Advanced Computing and Data Science (CACDS) and Research Computing Center (RCC) at the University of Houston. MSC thanks Yossi Eliaz for his intellectual discussion. AGG is supported by a fellowship from the Houston Area Biophysics Program (T32 GM008280). PWS acknowledges funding from Knut and Alice Wallenberg Foundation and the Swedish Research Council. MK acknowledges financial support from the Young Scholar Fund of the Universität Konstanz.

**Author contribution statement**

M.K., P.W.S., and M.S.C. designed research, F.Z., D.H., A.G., L.B., and M.K. performed research work and did data analysis. All authors contributed to discussions and data interpretation. M.S.C., P.W.S., M.K. D.H., F.Z., and A.G. wrote the manuscript.

**Conflict of interest statement**

There is no conflict of interest.

**References**

**TOC Graphic**

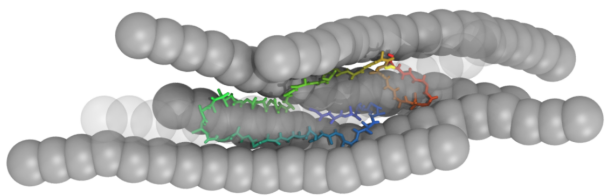

# Supplementary Information

## Crowding-induced Elongated Conformation of Urea-unfolded Apoazurin: Investigating the Role of Crowder Shape *In Silico*

Fabio C. Zegarra[&,1], Dirar Homouz[&,1,2,3], Andrei G. Gasic[1,3], Lucas Babel[1], Michael Kovermann[4,5*], Pernilla Wittung-Stafshede[6,*], Margaret S. Cheung[1,3,*]

& Equal contribution

1. Department of Physics, University of Houston, USA
2. Department of Physics, Khalifa University of Science and Technology, Abu Dhabi, UAE
3. Center for Theoretical Biological Physics, Rice University, USA
4. Department of Chemistry, Universität Konstanz, Germany
5. Research School Chemical Biology (KoRS-CB), Universität Konstanz, Germany
6. Biology and Biological Engineering Department, Chalmers University, Sweden

**Corresponding author, email:**
*Dr. Margaret Cheung mscheung@uh.edu
*Dr. Pernilla Wittung-Stafshede  pernilla.wittung@chalmers.se
*Dr. Michael Kovermann michael.kovermann@uni-konstanz.de

**1. The Hamiltonian of a protein at varying concentration of urea**

We adopted the "Urea-SC-DIH" model from a previous study [1] where solvent-mediated interactions $\varepsilon^{\alpha}_{ij}$ for a particular concentration of urea ($\alpha$) were used in the native side-chain interactions between the side-chain beads *i* and *j* and in the dihedral angle formed by four successive $C_\alpha$ beads.

$$E_p = E_{structural-urea} + E^{NB}_{ij\;SC-urea} + E^{NB}_{ij\;HB} + E^{REP}_{ij}, \quad (S1)$$

$$E_{structural-urea} = E_{bond} + E_{angle} + E_{dihedral-urea} + E_{chiral}. \quad (S2)$$

The potential energy of the protein ($E_p$) is the sum of a structural term ($E_{structural-urea}$) and several non-bonded terms. The list of native pair contacts between the *i*th and *j*th side-chain beads was established by the CSU software [2].

The structural potential term ($E_{structural-urea}$) includes a bond-length potential, bond-angle potential, dihedral-angle potential modulated by the concentration of urea, and a chiral potential term that considers the L-form of every side-chain [3]. The equilibrium values, used to estimate the potential terms, were taken from the crystal structure of apoazurin (PDB ID: 1E65) [4]. The disulfide bond between the side-chain beads of Cys3 and Cys26 was modeled by a bond-length potential.

The non-bonded potential term considers the native and non-native interactions among residues. The potential term for the native pairs consists of the side-chain-side-chain potential term ($E_{ij}^{NB}{}_{SC-urea}$) and the backbone hydrogen-bonding potential term ($E_{ij}^{NB}{}_{HB}$) [5]. The repulsive potential term ($E_{ij}^{REP}$) is regarded for interactions between $C_\alpha$ beads as well as between non-native side-chain pairs. The repulsive potential term is given by Equation (S3) where $\varepsilon = 0.6$ kcal/mol and $\sigma_{ij} = 0.9(\sigma_i + \sigma_j)$, $\sigma_i$ and $\sigma_j$ are the Van der Waals radii of $C_\alpha$ or side-chain beads.

$$E_{ij}^{REP} = \varepsilon \left(\frac{\sigma_{ij}}{r_{ij}}\right)^{12}. \tag{S3}$$

The influence of urea at a certain concentration was introduced in the Hamiltonian by using two terms as indicated by the protocol "Urea-SC-DIH" from our prior study [1]: the side-chain-side-chain potential term and the dihedral-angle potential term.

The side-chain-side-chain potential ($E_{ij}^{NB}{}_{SC-urea}$) for a specific concentration of urea (α) is given by a Lennard-Jones potential in Equation (S4) where the solvent-mediated interactions ($\varepsilon_{ij}^\alpha$) were derived from all-atom molecular dynamics simulations using a Boltzmann inversion approach [6]. α represents a solution with a concentration of urea of 0M, 6M, or 8M. In the absence of urea (i.e. 0M), the solvent-mediated interactions ($\varepsilon_{ij}^\alpha$) for the non-bonded interactions were obtained from the Betancourt–Thirumalai potential [7], in which case the solvent-mediated interaction between Thr and Thr residues was used to obtain the solvent-mediated interactions of other residue pairs. The distance between the side-chain beads from two different residues is represented by $r_{ij}$, and the equilibrium position between two beads is given by $\sigma_{ij} = 0.9(\sigma_i + \sigma_j)$, where $\sigma_i$ and $\sigma_j$ are the radii of the side-chain beads $i$ and $j$, respectively.

$$E_{ij}^{NB}{}_{SC-urea} = \varepsilon_{ij}^\alpha \left[\left(\frac{\sigma_{ij}}{r_{ij}}\right)^{12} - 2\left(\frac{\sigma_{ij}}{r_{ij}}\right)^6\right]. \tag{S4}$$

The dihedral-angle potential ($E_{dihedral-urea}$) for a specific concentration of urea (α) is given by Equation (S5) where the dihedral force constant depends on the average of the solvent-mediated interactions of four successive $C_a$ beads with indexes $i$, $j$, $k$, and $l$. In the absence the urea, $\bar{\varepsilon} = \varepsilon$.

$$E_{dihedral-urea} = \sum_{dihedrals}^{C_{a_i}-C_{a_j}-C_{a_k}-C_{a_l}} \bar{\varepsilon}[1 - \cos(\phi - \phi_0)] + 0.5\bar{\varepsilon}[1 - \cos(3(\phi - \phi_0))], \quad (S5)$$

where $\bar{\varepsilon}$ is given by

$$\bar{\varepsilon} = \left(\varepsilon_{ij}^\alpha + \varepsilon_{ik}^\alpha + \varepsilon_{il}^\alpha + \varepsilon_{jk}^\alpha + \varepsilon_{jl}^\alpha + \varepsilon_{kl}^\alpha\right)/6. \quad (S6)$$

The interaction between crowder and crowder ($E_{cc}$) of the spherical model of dextran 20 (DX20-S or DX20A-S) is given by a repulsive term as described by Equation (S3). $E_{cc}$ of a spherocylinder crowder (DX20-series or DX20A-series) is composed of the potential energy of crowders and the repulsive interaction between crowders. The potential term for the interaction between crowder and crowder is also described by Equation (S3), and the intra-potential energy of a spherocylinder crowder consists of a bond-length potential, bond-angle potential, and a dihedral-angle potential, which equilibrium values were set to the radius a crowder bead, 180°, and 0, respectively.

## 2. Results from the NMR measurements

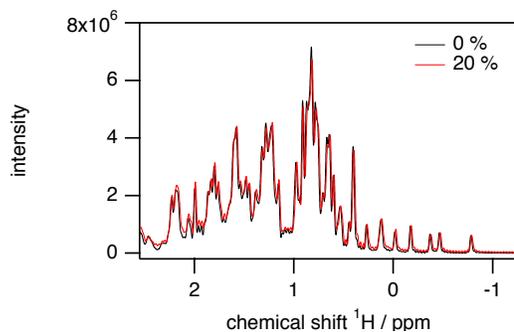

**Figure S1.** High-field chemical shifts for one-dimensional $^1$H NMR spectra for native apoazurin in absence (colored in black) and in presence of $c^{d20} = 200$ g/l (colored in red). Both spectra were recorded at $T = 298$ K and $B_0 = 20$ T.

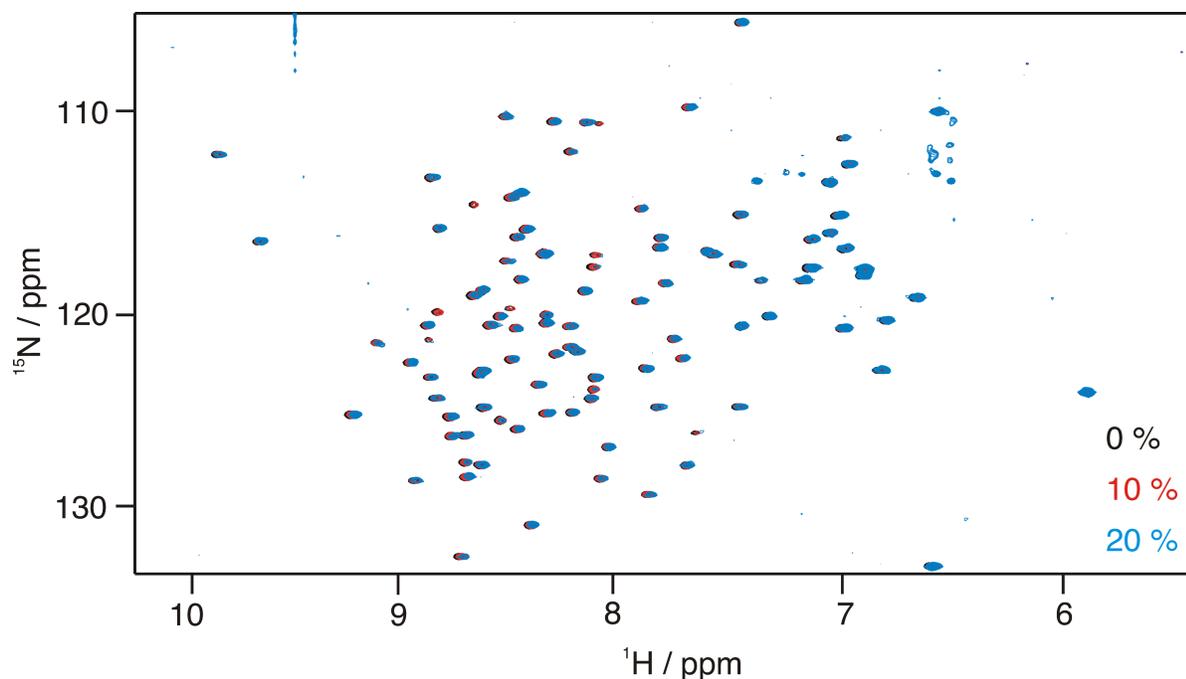

**Figure S2.** Two-dimensional $^1$H-$^{15}$N NMR HSQC spectra for native apoazurin at different concentrations $c$ of dextran 20 ($c^{d20}$ = 0 g/l colored in black, $c^{d20}$ = 100 g/l colored in red and $c^{d20}$ = 200 g/l colored in blue). All spectra were recorded for at $T$ = 298 K and $B_0$ = 20 T. The assignment of signals is found in [8].

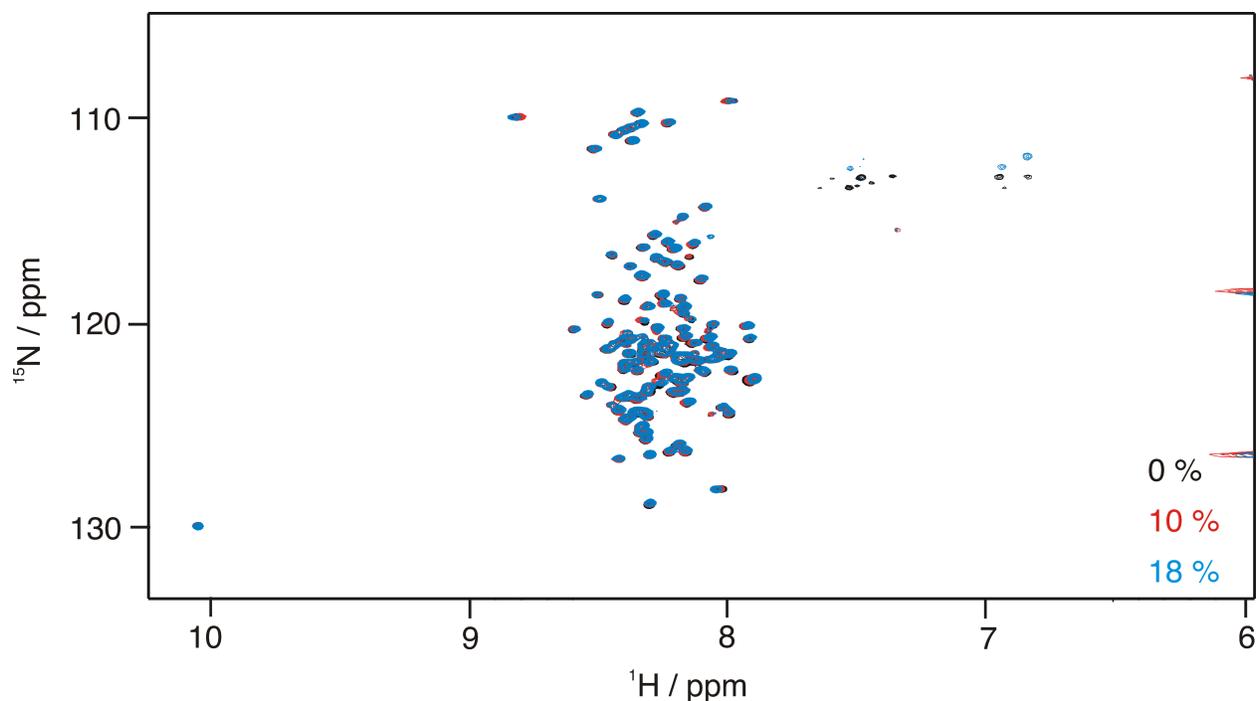

**Figure S3.** Two-dimensional $^1$H-$^{15}$N NMR HSQC spectra for denatured apoazurin ($c^{urea}$ = 6.8 M) at different concentrations $c$ of dextran 20 ($c^{d20}$ = 0 g/l colored in black, $c^{d20}$ = 100 g/l colored in red and $c^{d20}$ = 180 g/l colored in blue). All spectra were recorded at $T$ = 298 K and $B_0$ = 20 T.

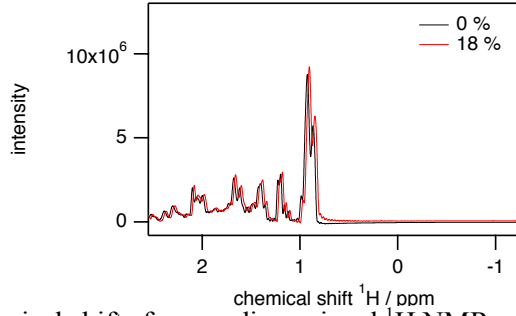

**Figure S4.** High-field chemical shifts for one-dimensional $^1$H NMR spectra for denatured apoazurin ($c^{urea}$ = 6.8 M) in absence (colored in black) and in presence of $c^{d20}$ = 180 g/l (colored in red). Both spectra were recorded at $T$ = 298 K and $B_0$ = 20 T.

## 3. The asphericity parameter Δ

The asphericity parameter $\Delta$ is defined as [9]

$$\Delta = \frac{3}{2}\frac{\left[\sum_{i=1}^{3}(\lambda_i-\bar{\lambda})^2\right]}{(tr\mathbf{T})^2}, \tag{S7}$$

where $\lambda_i$ is an eigenvalue of the inertia tensor $\mathbf{T}$ and $\bar{\lambda}$ is the average eigenvalue given by

$$\bar{\lambda} = \frac{tr\mathbf{T}}{3}, \tag{S8}$$

$tr$ is the trace of $\mathbf{T}$. The elements of $\mathbf{T}$ are obtained from the following expression,

$$T_{\alpha\beta} = \frac{1}{2N^2}\sum_{i,j=1}^{N}(r_{i\alpha}-r_{j\alpha})(r_{i\beta}-r_{j\beta}), \tag{S9}$$

where $\alpha$ and $\beta$ represent the axes $x$, $y$, and $z$. $N$ is the number of beads of the coarse-grained protein model, and $r_{i\alpha}$ is the component of the position of the bead $i$ in the $\alpha$-direction.

## 4. Analysis of $r_H$ and $R_g$

Hydrodynamic radius $r_H$ is an effective radius, which assumes the diffusing protein is spherical. However, as the asphericity $\Delta$ of the protein increases, $r_H$ and radius of gyration $R_g$ diverge. For pedagogical purposes, we use an extreme case to see how both $\Delta$ and $R_g$ are needed to describe $r_H$. This example assumes that as the unfolded protein becomes less spherical, it becomes more like a cylindrical rod with increasing aspect ratio $AR$. We can write an expression $r_H = R_g f(\Delta)$ by using the Stokes-Einstein relation

$$D_{sphere} = \frac{k_B T}{6\pi\eta r_H}, \tag{S10}$$

where $D_{sphere}$ is the diffusion coefficient of a sphere, $k_B$ is the Boltzmann constant, $T$ is temperature, and $\eta$ is the viscosity, and also using the modified relation for a rod [10],

$$D_{rod} = \frac{k_B T}{6\pi\eta r AR}\left(\ln(AR) + a + \frac{b}{AR} + \frac{c}{AR^2}\right), \tag{S11}$$

where $D_{rod}$ is the diffusion coefficient of a rod, and $a, b, c$ are constants ($a = 0.312$, $b = 0.565$, $c = 0.1$). Equating $D_{sphere}$ to $D_{rod}$, we can find the (effective) $r_H$ for the rod, which is

$$r_H = \frac{rAR}{\left(\ln(AR) + a + \frac{b}{AR} + \frac{c}{AR^2}\right)}. \tag{S12}$$

By expressing $AR$ as a function of $\Delta$

$$AR(\Delta) = \sqrt{\frac{2\Delta + 3\sqrt{\Delta} + 1}{1 - \Delta}}, \tag{S13}$$

and letting $R = R_g$, we can show how $r_H/R_g$ diverges as $\Delta$ of the protein increases (see Figure S5). Since $AR$ is the ratio of $z$ and $x$ principle components of gyration tensor and since $\Delta$ is a function of the principle components, we can obtain $AR(\Delta)$ by assuming $x$ and $y$ principle components are equal (as in a cylinder).

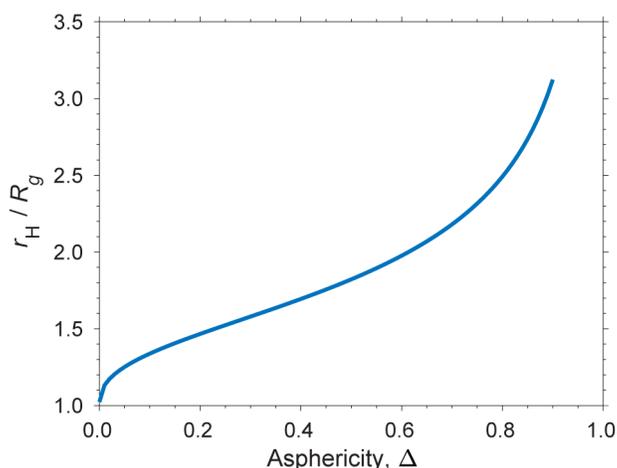

**Figure S5.** Hydrodynamic radius $r_H$ over radius of gyration $R_g$ of a protein with respect to asphericity $\Delta$.

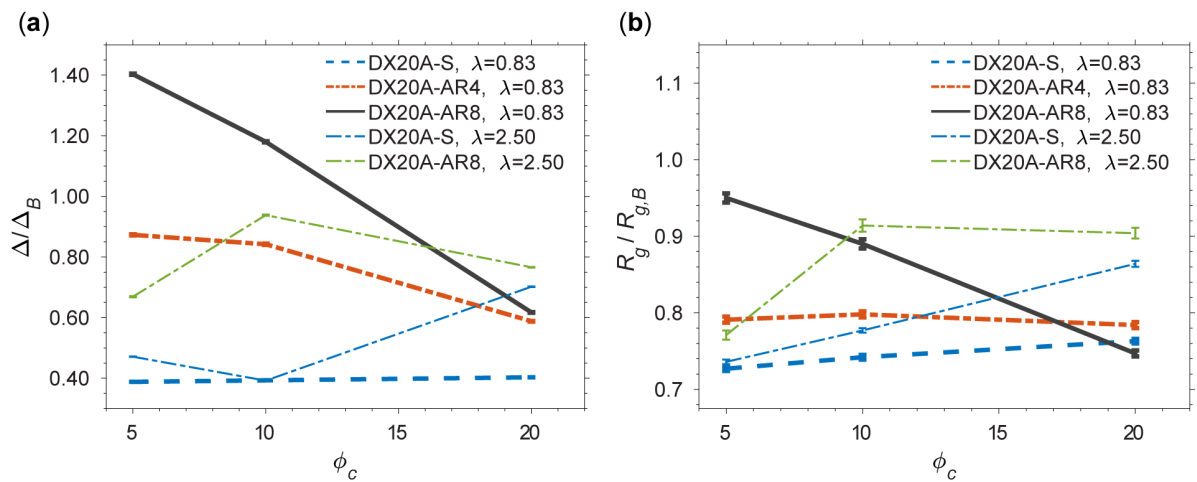

**Figure S6.** (a) Average asphericity (Δ) of protein apoazurin in varying shapes of crowders normalized by the average asphericity ($\Delta_B$) of protein apoazurin at the bulk condition against the volume fractions of crowders ($\phi_c$). (b) Average radius of gyration ($R_g$) of apoazurin normalized by the average radius of gyration under the bulk condition against $\phi_c$. The interactions between the protein apoazurin and crowders are attractive. The concentration of urea is 6 M. The temperature is $k_BT_u/\varepsilon$=1.14, corresponding to a room temperature (see Method), for all conditions. Error bars are included.

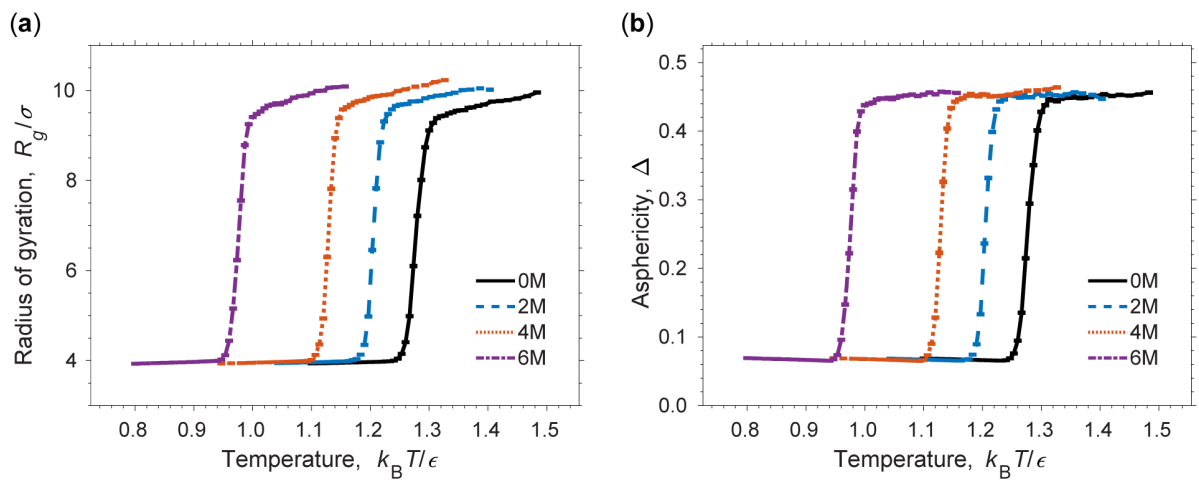

**Figure S7.** (a) The radius of gyration $R_g$ and (b) asphericity Δ of apoazurin as a function of temperature for apoazurin at several concentrations of urea (0M, 2M, 4M, 6M). Error bars are included.